\documentstyle[12pt,aaspp4]{article}
\def\etal{{\it et al.\/}}
\begin{document}

\title{\bf An Unusual Radio Galaxy in Abell 428: A Large, Powerful FR I Source In A 
Disk-Dominated Host} 

\author{Michael J. Ledlow$^{1,2}$}
\affil{New Mexico State University, Dept. of Astronomy \\ Las Cruces, NM 88003}
\vskip 0.2in
\author{Frazer N. Owen$^{1}$}
\affil{National Radio Astronomy Observatory$^3$ \\ 
Socorro, NM 87801}
\vskip 0.2in
\centerline{and}
\vskip 0.2in 
\author{William C. Keel$^1$}
\affil{University of Alabama, Dept. of Physics and Astronomy\\ 
Tuscaloosa, AL 35487}
\vskip 1.0in
$^1$Visiting Astronomer, Kitt Peak National Observatory, National Optical 
Astronomy Observatories, operated by the Association of Universities for 
Research in Astronomy, Inc., under contract with the National Science 
Foundation. 

$^2$Present Address: Institute for Astrophysics, University of New Mexico,
Albuquerque, NM 87131 

$^3$The National Radio
Astronomy Observatory is operated by  Associated Universities, Inc., under
contract with the National Science Foundation.

\clearpage

\begin{abstract} 

We report the discovery of a powerful ($\sim 10^{24}~{h_{75}^{-2}}~W~Hz^{-1}$ 
at 20cm) 
FR I radio source in a highly flattened disk-dominated galaxy.  
Half of the radio
flux from this source is concentrated within the host galaxy, with 
the remainder in a pair of nearly symmetrical lobes with 
total extent $\sim 200$ kpc nearly perpendicular to the disk.  The
traditional wisdom maintains that powerful, extended radio sources 
are found only in ellipticals or recent merger events. 
We report 
B, R, J, and K imaging, optical spectroscopy, a rotation curve, an IRAS detection, 
and a VLA 20 cm image for this galaxy, 0313-192.  The optical and near infrared
images clearly show a disk. 
We detect apparent spiral arms in a deep B band exposure, and a dust-lane from 
a higher resolution B band image. 
The reddened nucleus is consistent with
extinction by a similar dust-lane.  
The optical spectrum suggests a
central AGN and some evidence of a starburst, with both the AGN
and central starlight appearing substantially reddened (perhaps by the
optical dust-lane). 
From analysis of the extended line emission in [OIII] and
$H\alpha$, we derive a rotation curve consistent with an early-type, 
dusty spiral seen edge-on.  From the IRAS detection at 60 and 100$\mu m$
we 
find that the ratio of Far IR to radio flux places this object firmly 
as a radio galaxy (i.e. the radio emission is not powered by star formation). 
The radio structure suggests that the radio 
source in this galaxy is related to the same physical mechanisms present
in jet-fed powerful radio sources, and that such powerful, extended 
sources can (albeit extremely rarely) occur in a disk-dominated 
host. 

\end{abstract}

\clearpage 

\section{Introduction and Background}

     Current wisdom on the nature of powerful, extended radio sources 
reflects the
observation that whenever an unequivocal morphological classification 
is possible, the host galaxies are always ellipticals 
or recent merger remnants (Wilson \& Colbert 1995, Urry \& Padovani 
1995, Antonucci 1993). This trend has driven several explanations
for how the environment within the galaxy affects the large-scale
radio structure.
The difference between radio-loud and radio-quiet objects has recently 
been explained by Wilson \& Colbert (1995) as connected to different 
host galaxy populations. The merger of two radio-quiet
disk galaxies and the subsequent spin-up of a binary black hole produced 
in the merger event (assuming the individual holes are of approximately 
equal mass) would produce central objects of the greatest angular momentum 
uniquely in elliptical hosts or pre-elliptical mergers. 
Wholesale merging in clusters is implied by the high spiral fractions
seen in galaxy clusters at high redshift as compared to the present epoch, 
as spirals and S0's merge to form ellipticals over cosmic time.  
Hutchings, Janson, \& Neff
(1989) and Smith \etal\ (1986) have argued that the host galaxies of 
radio-quiet quasars are spirals whereas the radio-loud quasars are found 
in elliptical hosts, consistent with the distinction in parent populations. 
However, HST imaging ({\it e.g.\/} Bahcall et al. 1996) has shown that
reality is more complex; there exist unambiguous elliptical
hosts of several radio-quiet QSOs, and probable spiral hosts for some 
radio-loud objects. The radio emission in many of these objects comes strictly
from a small core rather than extensive double lobes, so it is not yet clear 
that the test is significant as regards to extensive double radio sources. 

     The standard picture of FR I and FR II radio sources (Fanaroff \& Riley 
1974) is that the extended structures in both classes are fueled by an active 
nucleus with outflows or jets that transport the radio plasma to large distances
from the nucleus. For radio powers $> 10^{23}~{h_{75}^{-2}}~W~Hz^{-1}$ at 20cm, 
nearly all radio sources fit into one of the two FR categories 
and are hosted by elliptical galaxies (or recognizable variations
on ellipticals) as found by Heckman et al. (1986) and Zirbel (1996, and references
therein) for radio sources not selected to reside in rich cluster environments.
In rich clusters the host galaxies appear to be much more homogeneous in nature, 
with a lower frequency of obvious merger activity (Ledlow \& Owen 1995b).  

    Powerful starbursts and spirals are frequently observed with 
radio luminosities $> 10^{22}~{h_{75}^{-2}}~W~Hz^{-1}$, but the origin of the 
radio emission seems to be very different from the FR jet-fed powerful radio 
galaxies. In both starbursts and luminous spirals (Hummel 1981, Condon \etal\ 
1982), the radio emission is almost always limited to the nucleus and disk, and 
is dominated by either non-thermal emission from cosmic rays produced in 
supernovae or thermal emission from HII regions (both of which closely trace the
distribution of star formation). 
Radio surveys of Seyfert galaxies (Ulvestad \& Wilson 1984a,b, 1989; 
Unger \etal\ 1986; Baum \etal\ 1993; Colbert \etal\ 1996a,b)
which are nearly always identified with spiral/S0 hosts, indicate that 
the radio emission may arise from 3 components: (1) sub-kpc emission 
associated with the nucleus; jets are frequently detected, but with no 
preferred 
axis relative to the galaxy disk (Baum \etal\ 1993). 
 (2) Extra-nuclear (kpc-scale) emission, 
and (3) $>$ kpc-scale emission associated with the disk or spiral 
arms of the galaxy.  The $>$ kpc-scale emission is most often diffuse, 
resembling {\it bubbles}, with no indication for tightly collimated outflow
(implied opening/cone angles are typically $60^\circ$).
The origin of the extra-nuclear emission is 
somewhat uncertain, although current models favor circumnuclear
starbursts in which the cosmic ray electrons are {\it blown out}
along the minor axis either trans/supersonically from pressure-driven winds 
(from either the starburst or an AGN) or subsonically in buoyant plumes
(Baum \etal\ 1993).  The correlation between the radio 
emission and starburst activity is additionally supported by the tight 
Far IR/Radio correlation observed in spiral galaxies and most Seyferts (Baum \etal\ 1993; 
Wilson \etal\ 1992; 
Rodriguez-Espinosa \etal\ 1987).  To the contrary, powerful radio galaxies 
always show significantly more radio emission than predicted by the FIR/Radio
correlation. 
In no case has extended radio emission been detected on $>$ 100 kpc scales 
in either luminous starburst or Seyfert galaxy samples.  The largest 
scale outflows detected in Seyfert galaxy samples span tens 
of kiloparsecs with little evidence for collimated emission on these scales 
(Colbert \etal\ 1996a; Kukula \etal\ 1995).  The radio emission in AGN 
spiral galaxies often resembles the features of classical radio galaxies, 
but at much lower radio powers and with sizes up to 2-3 orders of magnitude
smaller in scale. 

Additionally, the radio luminosity function of Seyferts, spirals, and radio 
galaxies are distinct and do not join one another, suggesting either 
differing parent populations or physical processes responsible for the radio
emission (Meurs \& Wilson 1984).  
It has been suggested that the parsec-scale jets which are seen in many Seyfert
and spiral galaxies are smothered or disrupted by the dense surrounding 
gas and turbulent velocity field in the circumnuclear environment and are
unable to escape to feed large-scale structures.
It seems surprising, however, that the {\it jet-smothering} processes should be 
100\% efficient, but to date, there have been only a few observed 
$>$kpc-scale collimated outflows from spiral or S0 hosts (Wilson 1996), and 
no confirmed spiral or disk-dominated galaxy has ever been 
identified with extensive double-lobed 
FR I or FR II-like radio morphology (V{\'e}ron \& V{\'e}ron-Cetty 1995).  

The issue of whether spiral galaxies can ever host powerful FR I or II radio
sources has a checkered history, exacerbated by morphological 
misclassifications 
and radio positions of insufficient accuracy. 
Shaver et al. (1983) investigated the
spiral galaxy associated with the FR II source PKS 0400-181, and found that 
this is most probably a superposition along the line of sight and that the
radio source is located at $z \sim 0.5$. The properties of this 
system were recently confirmed as a superposition of a spiral at z=0.0367
and a background elliptical at z=0.341 (R\"onnback \& Shaver 1996). 
Identification issues also cloud
the reported identifications of 3C 178 and 4CT 69.05.1 with nearby spirals. 
Further confusing the issue
is the appearance of spiral-like features in emission lines in otherwise
elliptical or S0 host galaxies, as in 3C 33 (Simkin \& Michel 1986), 3C 305 
(Heckman \etal\ 1985), 
PKS 1718-649 (Keel \& Windhorst 1991), and IRAS 0421+040 (Beichman \etal\ 1985; 
Hill \etal\ 1988; Holloway \etal\ 1996) among others. 

Where do S0 galaxies fit in the radio galaxy heirarchy?  In one of the
largest samples of S0's, Sadler \etal\ (1989) determined that the 
probability for radio emission was significantly lower than for 
ellipticals at the same total luminosity.  However, if one considered
the bulge component alone, the amplitude of the radio luminosity function
for S0's and ellipticals was nearly identical at the same optical luminosity. 
In the radio
S0's almost always resemble the structures seen in spiral galaxies
(compact emission limited to the nucleus or disk of the galaxy). 

Zirbel (1996) reported
6 possible S0? radio galaxy identifications from a large sample of 
very powerful radio galaxies ($10^{26} < P_{408MHz} < 10^{28}~h_{50}^{-2}~
W~Hz^{-1}$).  These candidate S0's were characterized by high central 
surface-brightness (a bright bulge component, typical of other D or cD galaxies
in her sample) and an outer profile which  
fell more steeply than an average elliptical.  However, none of the 
6 objects classified as S0? exhibited spiral features or possessed a
significant flattened disk component.  Much of the ambiguity associated with these 
classifications results from the heterogeneous nature of the S0 class.
We discuss this point further in section 4. 

To date, only four historically classified disk-like galaxies have been 
claimed to exhibit FR-like radio morphology
(collimated $>$kpc jets and lobes extending beyond the optical size of the
host galaxy); 
3C 305, 3C 293, NGC 5972, and NGC 0612. With the exception of NGC 5972
all of these objects are listed in 
a catalog of dusty ellipticals (Ebneter \& Balick 1985). Both 3C 293 and 3C 305 
are flattened ($\epsilon\sim 0.4-0.5$) and possess a disk of rotating 
emission-line gas (Heckman \etal\ 1985).  
However, Heckman \etal\ found no evidence for a stellar 
disk in either system (stellar rotational velocities significantly less than
the emission-line gas), and classified both objects as peculiar ellipticals. 
The radio source in 3C 305 is also quite small (a few kpc).  
From a sample of 347 nearby radio galaxies, NGC 5972 (classified as S0-a) 
was the only {\it disk} galaxy associated with a radio source significantly 
larger than
the optical galaxy (500 kpc; V{\'e}ron \& V{\'e}ron-Cetty 1995).  
Similarly, NGC 5972 possesses a rotating disk of high-ionization emission-line 
gas with no observable stellar disk, and was reclassified as an elliptical
by the authors.  NGC 0612 is perhaps the most enigmatic
of this group of objects.  Ekers \etal\ (1978) noted that both a disk and 
nuclear bulge were quite distinct.  The galaxy is bisected by a prominent
dust-lane, somewhat similar in appearance to the Sombrero galaxy (M104) or 
Centaurus A. 
However, recent optical surface photometry 
of this system revealed that the outer profile (ignoring the prominent 
inner dust-lane) follows an $r^{1/4}$-law
nearly perfectly (Fasano, Falomo, \& Scarpa 1996). 
Consistent with the other objects in this group, Goss \etal\ (1980) measured 
extended, rotating emission-line
gas out to 40 kpc from the nucleus. In all of these cases, a recent merger 
between an elliptical and a gas-rich disk-galaxy may be consistent with 
the observed properties and morphology, yet none of the four objects
are dominated by a significant disk component.  Thus to date there 
are no confirmed identifications of a large-scale, jet-fed radio source 
from a disk-dominated host, despite galaxy classifications as 
disk-like systems. 

   From our VLA 20cm survey of $>$ 500 Abell clusters 
(Ledlow \& Owen 1995a; Owen, White, \& Ge 1993; Owen, White, \& Burns 1992), 
we have found
one clearly disk-dominated galaxy with radio morphology resembling the powerful 
FR I radio sources.  The survey was statistically complete to 
$10^{23}~h_{75}^{-2}~W~Hz^{-1}$ 
out to z=0.09 though many sources were detected at lower redshifts with
radio luminosities below this cutoff, including a number of spiral/S0
galaxies. 

     Abell 428 (Richness Class 0, Revised Rood-Sastry type C8, Bautz-Morgan type 
III) is very unusual in that nearly half of the galaxies within the inner 
1 Mpc radius of Abell's cluster position appear to be disk galaxies. 
Two radio sources were identified with this cluster, and 
cluster membership was confirmed from optical spectra (Owen, Ledlow, \& 
Keel 1995).  0313-192 (z=0.067, the brighter of the two at $S_{1400MHz}=98$ mJy)
is located
0.05 Abell radii or $100h_{75}^{-1}$ kpc from Abell's 
nominal cluster center. The total angular extent of its radio lobes is 
$\sim 200h_{75}^{-1}$ kpc; nearly 100 times larger than the scale-height
of the disk observed nearly edge-on! 
The other radio source (0314-192) is $\sim 300h_{75}^{-1}$ kpc from 
the cluster center, is weaker (17 mJy), and is found in an elliptical 
host galaxy. R-band surface-photometry was reported for both sources 
in Ledlow \& Owen (1995b).  

In this paper, 
we present radio, optical, near-IR imaging, an IRAS detection, 
and optical spectroscopy of 
this galaxy. We examine the overall properties
of the host galaxy to 
investigate possible peculiarities which might 
account for a disk galaxy showing such an unusual radio source.
We present evidence which suggests that the radio emission in this
galaxy is consistent with large-scale nuclear outflows driven by an 
imbedded AGN.  From the combined analysis of the optical and near-IR imaging
and optical spectroscopy, we argue that this galaxy is most consistent with
an early-type spiral (Sa-Sb) classification. This appears to be the first 
detection of a powerful FR I radio galaxy with radio structure greater than
100 kpc in size in an unambiguously disk-dominated
host (whether S0 or genuine spiral).  
In section 2 we describe the observations.  In section 3 we discuss the
results from the radio, far IR, optical and near-IR imaging, 
and optical spectroscopy.  We present our interpretation of the 
nature of this object and our conclusions in the 
discussion of section 4.  

\section{Observations}

   We have obtained followup optical imaging of 0313-192 in A428 at B-band 
using the 
KPNO 2.1m telescope on 1 January 1995.
The KPNO $B$ image is a 
9-minute exposure obtained through thin cirrus with $1.0\arcsec$ seeing
using a 
1024x1024 Tektronix CCD at a scale of 0.30 arcsec/pixel.  
We also observed 0313-192 in $B$ using the Astrophysical Research 
Consortium's (ARC) 3.5m telescope at Apache Point Observatory (APO) on 
13 October 1996.  We obtained 13 10-min dithered exposures using a 
2048x2048 Tek/SITe CCD.  The chip was binned 2x2 on readout
to produce a plate-scale of 0.28 arcsec/pixel. 
Conditions were non-photometric, and the seeing was moderate at 
$\sim 1.3\arcsec$. 
We observed A428 in the Cousins $R$-band using the KPNO 0.9m telescope 
on 11 August 1995 with the T2KA Tektronix 2048x2048 CCD. 
The $R$ image has a total exposure time of 
30 minutes with $<1\arcsec$ seeing during photometric conditions with a 
scale
of 0.68 arcsec/pixel.  For comparison, all these images were interpolated
to a common grid with 0.3 arcsecond pixels.
Magnitudes 
were transformed to the standard system from zero-point and extinction 
calibration from observations of Landolt (1992) equatorial standard stars.  

    The near-infrared $J$ and $K$ images were obtained at the ARC 3.5m telescope 
on 8 December 1995 using 
the GRIM II 256x256 NICMOS array in F/5 mode (0.48 arcsec/pixel). 
Seeing at $J$ and $K$ was 0.8 arcsec and sky conditions were photometric.
Magnitudes were transformed to the standard system using observations 
of IR standard stars from Carter \& Meadows (1995).  At $J$, we obtained 
ten 60-second dithered exposures, alternating between source and sky. 
At $K$-band, we coadded 25 10-second exposures.  

     Following an initial blue spectrum for redshift measurement as part
of a broader survey of radio sources in Abell clusters
(Owen, Ledlow, \& Keel 1995), we obtained an optical spectrum 
covering the broader
range 3800--7400{\AA}.
This spectrum was obtained with the KPNO 2.1m telescope 
and GoldCam CCD spectrometer.
To reduce problems with atmospheric dispersion at this southerly
declination, 0313-192 was observed with the 2-arcsecond slit oriented 
north-south, which proved somewhat unfortunate since the disk is aligned
nearly east-west. The exposure time 
was 20 minutes at a dispersion of 2.5 {\AA}/pixel and resolution of 
$\sim 6$\AA\/. Flux calibration used 
observations of the spectral standard star Feige 54.  

     Additionally, we obtained a high-resolution spectrum of 0313-192
using the ARC 3.5m telescope on 12 October 1996.  We used a 2$\arcsec$
slit, rotating the instrument $\approx 6^{\circ}$ to place the slit
through the disk of the galaxy.  We used the Double-Imaging Spectrograph
(DIS), which separates the incoming beam
blueward and redward of 5550\AA\ using a dichroic filter.  The 
detectors are a Tektronix 512x512 CCD on the blue-side, a
TI 800x800 chip in the red.  The plate-scales are 1.086 and
0.61 arcsec/pixel on the blue and red chips respectively. 
We used the highest resolution 
gratings (830.8 lines/mm in the blue and 1200 lines/mm in the red) 
with dispersions of 1.6 (blue) and 1.3 (Red) \AA\ /pixel.  The grating 
tilts were adjusted to center redshifted [OIII]$\lambda 5007,4959$ and 
$H\beta$ on the blue chip, and 
$H\alpha$+[NII] in the red.  We obtained 5 30-min 
exposures. The data were calibrated separately and coadded in two
dimensions for analysis of possible extended line emission.  

     Radio observations were made at 1400 MHz with the VLA in both the 
B and C-arrays. The observations were made in snapshot mode (see Owen \& 
Ledlow 1997) with typical integrations of 6 minutes.  The data for each 
array were 
self-calibrated and combined in the UV-plane.  We applied the primary-beam
correction to the final map.  

\section{Results from Radio, Far-IR, Optical, and Near-IR Observations} 

\subsection{Radio Emission} 

    Figure 1 shows an overlay of the 20cm radio emission (contours) onto 
the optical $R$-band image (grey-scale).  Each of the radio 
lobes extend approximately 100 kpc in size; nearly two orders of magnitude
larger than the scale-height of the disk seen nearly edge-on. 
Drawing a bisector from SW to NE through the radio lobes and galaxy 
nucleus we 
find an orientation PA=$18^{\circ}$ (north through east).  The projected
major axis of
0313-192 is oriented at PA$\sim 84^{\circ}$ on the $R$-band image,
resulting in an inclination of the radio-lobe axis to the 
disk of $66^{\circ}$ in the plane of the sky.  In comparison with some
other reported identifications of extended sources in spiral
galaxies, we note that the optical and radio cores coincide
to within 1". We can rule out a superposition with an elliptical cluster
member of A428 given that we detect no trace of a  halo surrounding 
0313-192 down to very faint surface-brightness levels.  If a superposition
were responsible for our identification the host object would have to 
be at a much greater distance.  
The {\it a priori} probability of a chance 
superposition with a spiral galaxy with $m_B\leq 17.5$ to within a 
coincidence of 1" is $<$0.1\% for our entire VLA survey of more than 
500 clusters. 
(using the galaxy counts of Kirshner, Oemler, \& Schechter 1979). 

    The integrated radio flux density from 0313-192 is $\sim 98$ mJy. 
Nearly half of this (48 mJy) originates in the 
unresolved galaxy nucleus+disk.  The northern radio lobe has an integrated
flux density of 32 mJy, the southern lobe 18 mJy. 
Both radio lobes show indications of more diffuse emission 
out to larger distances, but deeper integrations will be necessary 
for a significant detection.  With the radio image presented in this paper
we do 
not detect actual {\it jet} emission ({\it i.e.\/}  Bridle \& Perley 1984). 
We have recently obtained followup higher resolution and higher frequency radio 
observations to look for evidence of collimated jet emission. 
We have confirmed the detection of a jet $\approx$ 36 kpc in length 
from the core extending into the southern lobe (Owen, Ledlow, \& Keel 1997). 

     Based on the radio morphology, we classify 0313-192 as a powerful 
FR I radio galaxy.  The morphology is very similar to the Fat-Double 
class defined in Owen \& Laing (1989).  The relationship between 
optical ($M_{24.5}(R)=-21.26$) and radio luminosity 
($\log P_{20cm}=23.95~W~Hz^{-1}$)
places 0313-192 very near the FR I/II break in the radio/optical luminosity 
plane (Ledlow \& Owen 1996).  The optical galaxy is one of the faintest
in our survey, as no confirmed radio identifications out to $z=0.09$ were
found with $M_{24.5}(R)$ fainter than -21.  

\subsection{IRAS Far-Infrared Emission}

     We used the XSCANPI utility (to coadd individual IRAS scans) to 
estimate IRAS fluxes or upper limits for 0313-192 at 12, 25, 60, and 
100$\mu m$.  
At 60$\mu m$ we found a good detection with a flux
density of $110\pm23~mJy$.  At 100$\mu m$ we measure a flux density of 
$470\pm 117$. 
There was no detection at either 12 or 25$\mu m$.  
Following Fullmer \& Lonsdale (1989), we define the FIR flux 
as : 
\begin{equation}
FIR = 1.26(2.58\times10^{12}f_{60}~+~1.00\times10^{12}f_{100})10^{-26}~W~m^{-2}
\end{equation}
where $f_{60}$ and $f_{100}$ are the IRAS flux-densities at 60 and 100$\mu m$. 
Using the flux densities listed above, we find $FIR=9.50\times10^{-15}~W~m^{-2}$. 

    The correlation between FIR and radio emission for spiral galaxies and 
Seyferts can be quantified via the parameter $q$ (Helou \etal\ 1985) as
\begin{equation}
q~=~\log{\Bigl(}{{FIR}\over{3.75\times10^{12}~W~m^{-2}}}{\Bigr)}~-~\log{\Bigl(}{
{S_\nu}\over{W~m^{-2}~Hz^{-1}}}{\Bigr)}
\end{equation}
where $S_\nu$ is the radio flux density.  Using the total integrated radio 
emission from 0313-192 (98 mJy) we find $q=0.41$.  Using only the radio emission 
associated with the nucleus+disk (48.5 mJy), we find $q=0.72$.  Both values
are considerably less than the canonical value of $<q>=2.27\pm0.20$ at 
$\nu=1.45~GHz$ for rich cluster spirals 
(Andersen \& Owen 1995) or $<q>=2.34\pm0.26$ from the entire IRAS bright galaxy 
sample 
(Condon 1991).  Agreement with these q-values would require a FIR flux 
35-75 times greater than is observed.  
In fact, many
powerful radio galaxies (both elliptical and S0) have strong FIR emission
(Golombek, Miley, \& Neugebauer 1988).  
The excess radio emission over that 
predicted by the FIR/Radio correlation confirms our suggestion that
this object be classified as a true radio galaxy, and provides no 
constraint as to the actual morphological type of this galaxy.  
Note that the spiral galaxy in the NW of the image in 
Figure 3 is 
only 39 arcsec ($46h_{75}^{-1}$ kpc) in projected distance from 0313-192.  
Given 
the low resolution of the IRAS satellite, we cannot rule out that 
some fraction of the FIR flux arises from this source. Thus the values
derived here should be viewed as strict upper limits.

\subsection{Optical/Near-IR Imaging and Isophotal Surface Photometry}

     In Figure 2, we show grey-scale images of 0313-192 in $B$(KPNO), $R$, $J$, 
\& $K$.  In Figure 3, we also show the deeper 
APO $B$-band image using two transfer functions; one to emphasize the 
blue properties of the galaxy disk, the other to examine the 
low surface brightness large-scale properties (down to 25.1 $\rm mag~arcsec^{-2}$ in the rest-frame of the galaxy). 
Inspection of the B-band images indicates several interesting features.
(1) From the deep B-image, extending both east and west from the nucleus 
are features that we identify with a spiral arm. Although the disk is viewed
nearly edge-on, the features clearly show an S-shaped symmetry about 
the nucleus.  The arm emission is asymmetric,
with the western arm somewhat brighter. 
(2)  There is no evidence to suggest a tidal interaction
with the galaxies to the SW or NW. There is an unusual feature at the
west edge of the disk (an extension to the south), 
but this features does not appear to have any 
connection to the neighboring galaxies. 
(3) From the shallower (but with better seeing) $B$-image in Figure 2, we see 
that the nucleus is 
diffuse, and the central brightness peak is bisected by a dark feature which 
we interpret as a dust-lane. 
The dust-lane has an apparent projected width of about 1 kpc (at the limit of 
our resolution).  Dust lanes are not uncommon features in radio galaxies, 
even among those classified as ellipticals.  De Koff \etal\ (1996) 
found that at least 30\% of powerful 3C radio galaxies show indications 
of dust, either in clear dust-lanes or chaotically distributed throughout
the galaxy.  Thus, the presence of a dust-lane does not confirm a 
spiral classification for 0313-192.  The spiral arm features present 
in the deep $B$ image provide much more substantial evidence for such 
a classification.  However, it is clear that the galaxy is strongly dominated by a 
flattened disk. 

    The nuclear continuum properties change strongly with wavelength. 
At $B$-band the nucleus is very diffuse with no obvious image core. 
Moving to redder wavelengths, even at $R$, the nucleus and/or bulge contributes
a substantial fraction of the galaxy light.  At $J$ and $K$ the emission is 
strongly dominated by an unresolved core (or bulge at the limit of 
our resolution; 1.1 h$_{75}^{-1}$ kpc).  We are able to trace the disk out to 
$\approx$ 20 kpc in radius along the major-axis.  Within a 40 kpc diameter aperture,
we calculate absolute magnitudes of -19.9, -21.4, -23.1, and -24.4 for 
$B$, $R$, $J$, and $K$ respectively (corrected for galactic extinction). 

We fit elliptical isophotes to each of the images using the 
method described in Jedrzejewski (1987).  The centroid, ellipticity, 
and isophotal position-angle were allowed to vary at each radius. 
We show the major-axis surface-brightness, ellipticity, and position-angle
profiles in Figure 4 for each color.  The data points were binned 
in order to sample the seeing disk at two points per resolution element. 
The ellipticity profiles are reasonably consistent at all colors, with 
a strong trend for $\epsilon$ to increase with radius up to a maximum 
$\epsilon \sim 0.75$. 
The outer optical isophotes are both highly elongated and ``pointy'',
indicating a highly inclined disk (since no true ellipticals have
ellipticity greater than 0.7). 
There are small variations in position angle between the 
optical and near-IR bands, most likely  related to the dust-lane or 
spiral features which are more prominent at the bluer colors. 

     In order to characterize the surface-brightness profiles for 
0313-192, for the $B$ and $R$-band images, we have fit a multi-component 
galaxy profile model of 
the form
\begin{equation}
\rm I(r) = S_b~exp(-7.668*[({{r}\over{r_e}})^{1/4}~-~1])~+ 
S_d~exp(-{{r}\over{r_0}}~+~({{r_1}\over{r}})^3)
\end{equation}
where $S_b$ and $S_d$ are the normalization scale factors for the 
bulge and disk components, $r_e$ is the de Vaucouleurs effective
(half-light) radius, and $r_0$ is the radius of the exponential 
disk.  
In Figure 5, we show the result of the best-fit models.  
For purposes of these fits we have excluded points within the 
inner two seeing radii.  
We find $r_e$ = 4.7, 1.9 kpc and $r_0$ = 6.6, 5.8 kpc for $B$ and $R$-band
respectively.  
The $r_1$-term was effectively zero. 
Neither profile is well-fit by a single-model (either $r^{1/4}$ or exponential
disk) over the entire range in radius.  
A pure disk model fits the profile quite well with the exception 
of the inner few kpc.  A pure bulge/$r^{1/4}$ model decays much too 
quickly to fit the profile beyond the inner several kpc. 
The larger $r_e$ observed at $B$-band is consistent with the diffuse core 
as compared to the $R$-band image.  For either value of $r_e$
our resolution is sufficient to resolve the spheroidal 
component.  Given the nature of the radio source, we expect that some
fraction of the nuclear/spheroidal component arises from an unresolved
core and AGN activity (as also evidenced by the very bright core at $K$-band). 
Thus, the bulge-to-disk ratio we measure should
be viewed as an upper-limit, which most likely includes a non-thermal 
component from an imbedded AGN. 

Following the method of Simien \& de Vaucouleurs (1986), 
we have the calculated the bulge-to-disk ratio for 0313-192 from the 
multicomponent fit.  Defining $k_1$ as the ratio of the integrated 
$B$-band luminosity of the spheroidal component to the total luminosity 
of the galaxy (both functions integrated to infinity), the bulge-to-disk
ratio (B/D) is $\gamma=k_1/(1-k_1)$.  Simien \& de Vaulcouleurs examined 
the dependence of the magnitude difference between the bulge component
and the total $B$ magnitude ($\Delta m=-2.5log(k_1)$) as a function 
of morphological type (T). 
We find $k_1=0.27$ and $\gamma=0.37$.  Using their 
interpolating functions for $\Delta m(T)$, we find T=2.6, consistent
with a Hubble classification of Sa-Sb.  As a comparison, S0 galaxies
almost always have $\gamma\geq 0.7$ (Dressler \& Sandage 1983; Bothun \&
Gregg 1990), with most $\geq 1$. 
0313-192
appears to fall in a region separate from S0's or lenticulars (T=-3-0) 
and ellipticals
(T$<$-3).  

Can we clearly rule out a S0 classification for 0313-192 based on 
the B/D ratio?  There is significant scatter in the $\Delta m(T)$ 
relation given by Simien \& de Vaucouleurs, which they claim is 
due to photometric and decomposition errors with little contribution 
from classification errors or cosmic scatter (however, see Kennicutt 1981; 
Sandage 1961).  
In the mean, however, 
S0's clearly have larger B/D ratios.  
More importantly, S0's are usually defined as disk galaxies 
which have no 
recognizable spiral structure (Bothun \& Gregg 1990).  Given the 
appearance of what appear to be spiral arms and the 
low B/D ratio, we believe that the most consistent classification 
for 0313-192 is that of an early-type spiral (Sa-Sb). 

\subsection{Color Profiles} 

     From the surface-brightness profiles at each color, we 
have calculated color profiles for $B-R$, $B-J$, $B-K$, $R-J$, $R-K$, 
and $J-K$.  These profiles are shown in Figure 6.  
We have corrected the individual colors for galactic extinction,
but have not applied a K-correction.  Errors were determined 
from RMS deviations of the magnitude within each isophote. 
We have included a 5\% calibration error for the near-IR bands,
and a 2\% uncertainty at $B$ and $R$. 
The inner few kpc of 
the galaxy is highly reddened relative to the color of the disk. 
The nucleus is $\approx$ 0.5-1 magnitude redder than
the disk depending on the particular color. 
In $R-J$ and $B-J$ the 
slope is much shallower from 2-4 kpc than in the other color
profiles.  At $J$-band, the core/bulge is much more promininent
than at $K$, at which the galaxy is dominated by an unresolved core. 

    All of the derived nuclear colors strongly suggest the 
presence of significant amounts of dust, either associated with the
nucleus itself or in the disk and associated dust-lane seen almost edge-on.  
Peletier \& Balcells (1996) derived color profiles for 45
early-type spiral galaxies from the Uppsala catalog to study colors
and color gradients in spiral bulges.  Excluding those galaxies
which they define as severely contaminated by dust, the average
nuclear $B-R$ color is near 1.5 (we observe $(B-R)_{nuc}\sim 1.9$).  
In addition, very few of their 
observed galaxies show strong gradients ($>$0.5 mag) from this 
value at increasing radius in the disk. With the exception 
of the redder $B-R$ color within the inner few kpc, 
0313-192 is very similar to the objects in their sample. 
In near-IR colors, 
typical nuclear $J-K$ values for ellipticals are $\sim 1$, for spiral
galaxies $J-K\sim 1-1.5$, and nuclear starbursts have $J-K$ in the 
range from 1-2 (Bothun 1991, Watson \& Gallagher 1997).  Watson 
\& Gallagher show a few examples of nuclear starbursts with $J-K$
$\geq 2$, although these are extreme cases.  
The nuclear $J-K$ color observed in 0313-192 is more typical 
of some Seyfert 1 galaxies and Quasars (Ward \etal\ 1982; Hyland
\& Allen 1982), although a significant amount of dust ($A_V > 5$)
could shift a weak starburst/AGN or normal early-type spiral 
to our observed value.  Additionally, our observed $B-K$ nuclear color 
is strongly suggestive of dust (or a significant non-thermal K-band flux).  
Typical colors
for spiral or S0 disks range from $B-K$=3-4.3 (Bothun \& Gregg 1990). 
We observe a nuclear $B-K$ color of $\sim 6$ although the disk 
appears consistent with standard values. 
We discuss the possible interpretations
of these results in the next section. 

\subsection{Nuclear Optical Spectra}

     In Figure 7, we show the KPNO optical spectrum for 0313-192. 
The y-axis is in units of $F(\lambda)$ in $ergs~cm^{-2}~sec^{-1}~
\AA^{-1}$.  
The 2-d spectrum was convolved with a $1\arcsec$ Gaussian along 
the spatial axis      
before extraction.  We also convolved the spectrum with a $5\arcsec$
Gaussian and extracted the spectrum separately. 
As the slit was oriented North-South, the
spectrum should have very little contribution from the underlying
disk of the galaxy.  

     The most prominent features in the spectrum are narrow
emission from [OIII]${\lambda}5007,4959$, H$\alpha$+[NII], 
[OI]${\lambda}6300$, and  weak [OII]${\lambda}3727$, 
H$\beta$, and [SII]${\lambda}6716+6731$. 
The narrow lines are suggestive of a Seyfert 2-like
spectrum or possibly a powerful starburst.  

     From the calibrated spectrum we have calculated line and 
continuum ratios and equivalent widths (EW) for 0313-192.  
These values are listed in Table 1. 
Of particular interest is the very high $H\alpha/H\beta$ ratio. 
We have corrected both $H\alpha$ and $H\beta$ for underlying 
stellar absorption, using the stellar absorption appropriate to
old bulge populations (Keel 1983). These corrections are 1.8 \AA \ for 
H$\alpha$ and 1.3
\AA\ at H$\beta$ (the correction at H$\beta$ is especially 
sensitive to any contribution from reddened hot stars).
Consistent with the color profiles, 
such a high ratio is indicative of substantial dust.  We do 
not detect any higher Balmer-line transitions in the spectrum. 
Another indication of the reddened core is in the 4000\AA\/ break
(D4000).  We find D(4000)=$2.15\pm0.18$, which is consistent
with the nominal value for elliptical galaxies, but much redder
than expected for early-type spirals or S0's (typical values
are from 1.7-1.9; Kennicutt 1992).  We measured all D(4000) values from 
the $F(\nu)$-calibrated spectrum.  From the larger convolution (which 
includes more of the off-nuclear light) we find a bluer break 
ratio ($1.96\pm0.14$) which may contain some contribution from 
the blue spiral-arms. 

     These line ratios are useful diagnostics of the relative 
importance of star-formation to a hard photoionizing continuum
indicative of AGN activity and are relatively insensitive to 
internal extinction, as discussed by Baldwin, Phillips, \& Terlevich (1981) 
and Veilleux \& Osterbrock (1987).  
These values place 0313-192 as an intermediate case between AGN 
and powerful starburst activity.  The [O I] strength is perhaps
the strongest diagnostic, since star-forming nuclei are not
observed with [O I]/H$\alpha$ as high as the 0.1 we find.
Based on the $[OIII]/[OII]/[OI]$ ratios
and fairly weak $H\alpha$ (which is an indicator of massive
star formation in addition to whatever nonstellar ionization is taking place), 
we suggest that the nuclear emission is dominated by 
weak AGN activity rather than a circumnuclear starburst. The substantial
reddening implied by the broadband colors and Balmer decrement might make
the nucleus intrinsically luminous, but seen from an unfavorable direction.

\subsection{Extended Line Emission and Rotation Curve} 

While the core spectrum is useful in illuminating the AGN or nuclear
starburst properties, off-nuclear spectra are required to search for 
evidence of a stellar disk.  The primary diagnostics
here are the $H\alpha$ equivalent width, the $H\alpha$/[NII] ratio, 
and the Tully-Fisher relation. 
While 
there is some overlap along the Hubble sequence in these properties
and S0's typically exhibit much more scatter than spirals, we can use
these diagnostics to estimate the probable morphological classification.
The largest uncertainty in these measurements is the unknown 
intrinsic extinction within the galaxy. Line-ratios are less sensitive
to extinction. 

From the high-resolution APO spectra centered on [OIII] and $H\alpha$+[NII],
we find the emission to be highly extended throughout the disk.  In 
Figure 8, we show rotation curves measured from these lines.  It is 
interesting to note that the gas distribution appears asymmetric, 
as we trace extended line emission out to 15 kpc west of the 
nucleus and only $\approx$ 10 kpc to the east.  This asymmetry is 
apparent in the $B$-band image of Figure 3, as the feature which we interpret
as a spiral arm is significantly fainter to the east.  The rotation 
curves are consistent with an edge-on dusty disk, with apparent 
solid-body rotation from the outer arms.  There is some indication that 
the rotation curve begins to level-off at $\approx$ 10 kpc in radius. 
We find good agreement between the [OIII] and $H\alpha$ velocities, although
the [OIII] luminosity clearly diminishes more rapidly with increasing radius 
(is more confined to the nuclear region). 
The [NII]/H$\alpha$ ratio in the disk is consistent with star formation 
([NII] much weaker than $H\alpha$) and HII-region like spectra. 
We adopt a 
maximum rotational velocity of $v_{rot}$=208 km sec$^{-1}$. 

\subsection{The $B$-Band Tully-Fisher Relation} 

Fouqu{\'e} et al. (1990) used a sample of 178 spiral galaxies in the 
Virgo cluster to determine the distance to the Virgo cluster against 
a calibrated B-band Tully-Fisher (TF) relation 
for 18 calibrator sources 
(with independent distance measurements.). 
Following
their method, we determine $B_T^C$ for 0313-192 measured to the 
25 mag arcsec$^{-2}$ isophote in the rest-frame 
of the galaxy (includes a correction for galactic extinction 
and $(1+z)^4$ surface-brightness diminution term). The largest
uncertainty arises from the correction for internal extinction. 
We estimate the inclination from the classical formula (Hubble 1926):
\begin{equation}
cos^2~i = {{q^2 - q_0^2}\over{1 - q_0^2}}
\end{equation}
where $q$ is the observed axis ratio (${{b}\over{a}}=1-\epsilon$)
measured at the 25 mag arcsec$^{-2}$
isophote and $q_0$ is the intrinsic axis ratio.  We estimate $q_0$ from 
the morphological type (T=2.6, from the bulge-to-disk ratio) using
the fit given in Fouqu{\'e} et al.  We find $q=q_0=0.27$ consistent with 
$i\sim 90^{\circ}$ (within an error of order $5^{\circ}$). 
For inclinations $>80^{\circ}$, Fouqu{\'e} et al. assume an internal 
extinction of 0.96 mags. Following a similar prescription, we find 
$B_T^C$=-20.79 (Fouqu{\'e} et al. use $H_0=68$) compared to -20.76 
predicted by the 
calibrated $B$-band TF relation.  Such strong agreement 
with the prediction suggests that 0313-192 is indeed rotationally supported.
Compared to the quartile ranges in 
the TF relation for the different spiral types, 0313-192 
falls cleanly within the Sa-Sb range, consistent with our previous 
classification.  Note, however, that there are claims that 
morphological segregation in the TF relation is an artifact of 
selection biases (Bottinelli \etal\ 1986; Fouqu{\'e} \etal\ 1990).  

Does the very good agreement with the TF relation rule out 
an S0 classification for 0313-192?  Dressler \& Sandage (1983) 
examined rotation curves for $\approx$ 30 field S0's.  Plotted on 
the TF relation S0's span the entire range of all spiral types. 
Dressler \& Sandage offer several possible explanations for the observed
scatter. They concluded that variations in the M/L ratio and/or significant 
kinetic energy in velocity dispersion (pressure support) between different
galaxies may explain the scatter.  S0's are thus a very heterogeneous 
class.  The distinction between an S0 
or early-type spiral on the TF relation varies significantly
from object to object.  The very good agreement for 0313-192 
with the TF prediction is strongly suggestive of an early-spiral 
classification, however we can not rule out a coincidental result. 

\subsection{$H\alpha$ Line Strength and Line Ratios} 

From the $H\alpha$ luminosity and equivalent width we compare 0313-192
to samples of early to late-type spirals and S0's.  
The EW measures listed in Table 1 corresond to nuclear values with
little contribution from the disk.  We use our high-resolution 
spectrum to measure the $H\alpha$ luminosity for disk and nuclear
components separately. 
After correcting the $H\alpha$ equivalent widths for underlying 
stellar absorption (1.8\AA\, see previous section), we find the
following values: integrated along slit (includes nucleus) EW=11.4\AA\ ,
pure disk (average of E and W components) EW=12.4{\AA}.  We found EW=9.6\AA\ 
from the nuclear spectrum 
in Figure 7.  Additionally, the disk has an emission spectrum 
consistent with HII regions ([NII] much weaker than $H\alpha$) whereas 
the ionization paremeter and line ratios in the nucleus are 
intermediate between a starburst and weak AGN activity.  We have
not corrected the $H\alpha$ fluxes for internal extinction. 

Kennicutt (1992) gives values for integrated 
$H\alpha$ EW for early Hubble-type spirals (omitting
nuclear starbursts and AGN) and found Sa=3, Sab=4.5, 
Sb=18, and Sbc=31\AA\/.  No corrections were made for 
internal extinction.  Based on these values, 0313-192 is placed firmly 
in the  Sab-Sb range.  For S0's, Pogge \& Eskridge (PE) (1993) observed
32 S0 galaxies selected by high {\sl HI} mass and chosen to be nearly 
face-on.  Both of these would enhance the observed equivalent
widths as compared to a more random sample, and likely as compared to 
0313-192 as well.  This sample therefore represents the upper envelope
of star formation in S0 galaxies. 14/32 of the objects were detected in 
$H\alpha$, 
with a large range in EW measures; they find the median $H\alpha$ EW=
$\rm 5.6\pm 2.4$ 
(median L($H\alpha$)=$2.51\times 10^{40}$ ergs sec$^{-1}$). 
The difference in the median L($H\alpha$) between 
the {\sl HI}-rich S0 sample and 125 early to late-type sprial galaxies 
from Kennicutt \& Kent 
(1983) (median L($H\alpha$)=$7.41\times 10^{40}$ ergs sec$^{-1}$) 
was statistically significant at the $>$99\% level. 
In comparison, we find an integrated L($H\alpha$)=$8.38\times
10^{40}$ ergs sec$^{-1}$ for 0313-192. 
Only 3/32 (9\%) of the S0's in the PE sample have integrated 
$H\alpha$ luminosites as high as 0313-192, and two of these objects
are classified as nuclear starbursts.  If we eliminate the two 
nuclear starburst galaxies, that leaves only 1/32 (3\%) S0's  
with comparable 
$H\alpha$ emission.  Statistically, 0313-192 is much more consistent
with an early-type spiral classification. 

\section{Discussion and Conclusions}

   The optical morphology of 0313-192 
is quite unusual for a source of this kind, with several indications 
that this is a spiral (and certainly a disk galaxy) rather than the
normal elliptical.  
The radio power and morphology of 0313-192 are typical of a large, luminous
FR I source, with the usual double-lobe structure and significant
core emission.  The radio lobes extend $\approx$ 100 kpc north and 
south of the disk. 
The nucleus is highly reddened, presumably from 
significant amounts of dust consistent with the observed dust-lane
bisecting the nucleus in the disk seen nearly edge-on.  
The nuclear optical and IR colors are consistent with 
either a highly reddened ($A_V > 5$ mag) core in a normal/starbursting 
spiral or from a (possibly non-thermal) quasar-like nucleus.  The very 
bright, dominant core at K-band supports a possible non-thermal 
component, or an intrinsically luminous core which is mostly obscured
at shorter wavelengths. 
The nuclear optical emission-line spectrum is intermediate between 
a starburst and weak AGN activity.  In particular, the [OI]/$H\alpha$
ratio (0.1) is significantly higher than found in circumnuclear starbursts.
The near-IR color and weakness of
FIR emission all argue in favor of an obscured AGN as the dominant
energy source. We also find no evidence for a galaxy merger or interactions 
between 0313-192 and the neighboring galaxies to the NW and SW.  This 
point is in stark contrast to the peculiar nature of all previous
radio galaxy candidates discussed in the introduction. 

    While all diagnostics point towards an early spiral type (Sa-Sb),
we cannot rule out an S0 classification given the large scatter 
and inhomogeneity of this class.  
Despite best efforts
to bin galaxies into morpholgical classes which represent a physically
meaningful stage of evolution or a homogeneity of properties, the 
true definition of an S0 remains somewhat mysterious.  
van den Bergh (1990) argues that ``the S0 classification type comprises
a number of physically quite distinct types of objects that exhibit
only superficial morphological similarities.''  This observation 
is consistent with the large scatter in fundamental properties such as 
the TF relation, central velocity dispersion, B/D ratio, the amount 
of gas and dust, and the degree
of flattening.  Thus, on an individual object basis, S0's may have 
arrived at their present morphology via very different evolutionary 
paths (van den Bergh 1997).  Additional confusion arises from the
multitude of different classification schemes. A particularly good 
example is the galaxy IC 310 in the Perseus cluster.  In the de Vaucouleurs
(1959)
classification scheme this galaxy was classified as SA(r)0, ({\it r} meaning
ring-like structure), an S0 in the Hubble/Sandage system, and a D galaxy 
in the Morgan (1962) form-family based on its extended halo. Yet by all
appearances IC 310 is a very 
round $L^{*}$-like elliptical ($\epsilon\sim 0.05$) with very faint, possibly 
shell-like features at large radii. Such faint peculiar features are not 
uncommon in powerful radio galaxies, and it is precisely these 
features (tidal tails, fans, shells, bridges) which support the contention
that galaxy mergers may be important in stimulating nuclear activity
(Heckman \etal\ 1986). 
Nearly all of the well-studied radio-loud 
{\it S0's} are consistent with an elliptical-like host which has 
undergone a merger event with a disk galaxy thus producing a weak
disk-component with significant dust.  

    Would an S0 classification 
diminish the apparent uniqueness of this object?  
We think not given 
that, regardless of the morphological classification, to the best of our 
knowledge no galaxy with the clearly disk-dominated features and apparent
spiral structure of 
0313-192 has ever 
been identified with a radio source of such large extent. 
As nearly half of the radio luminosity from this source arises in the 
lobes $\approx$ 100 kpc from the nuclear source, the origin of the radio 
emission is without question an AGN-driven, large-scale nuclear outflow 
rather 
than radio emission powered by star formation.  
In all respects, 0313-192 is consistent with the standard picture of 
powerful FR I radio galaxies, with the exception of its unusual optical 
morphology.  To the best of our knowledge, this is the first reported 
detection of a large-scale FR I
radio source in an unambiguously disk-dominated host galaxy. 

\acknowledgements

\noindent 
{\bf Acknowledgements}

M.J.L. thanks Nick Devereux and Alan Watson for helpful discussions 
on the FIR and NIR properties of spiral galaxies.
We thank Greg Bothun for his comments on an earlier draft of this 
paper. 
This work was partially supported by NSF Grant AST-9317596 to J.O. Burns
and C. Loken.     
This research has made use of the NASA/IPAC Extragalactic database (NED)
which is operated by the Jet Propulsion Laboratory, Caltech, under 
contract with the National Aeronautics and Space Administration.

\clearpage

\begin{deluxetable}{cccc}
\tablewidth{300pt}
\tablenum{1}
\tablecaption{Nuclear Emission-Line Strengths and Line/Continuum Ratios} 
\tablehead{
\colhead{Feature}   & \colhead{Flux} & 
\colhead{EW} & \colhead{Ratio} \\
\colhead{} & 
\colhead{$ergs~cm^{-2}~s^{-1}$} & \colhead{$\AA$} & 
\colhead{}}
\startdata
$H\alpha$&7.7e-15&9.6&\nodata \nl
$[NII]\lambda6586$&1.3e-15&2.0&\nodata \nl
$H\beta$&2.2e-15&2.8&\nodata \nl
$[OIII]\lambda5007$&1.8e-14&19.6&\nodata \nl
$[OII]\lambda3727$&9.7e-16&3.3&\nodata \nl
$[OI]\lambda6300$&6.9e-16&0.6&\nodata \nl
$[SII]\lambda6716+6731$&1.2e-15&0.5&\nodata \nl
$[OIII]/H\beta$&\nodata&\nodata&7.9 \nl
$[NII]/H\alpha$&\nodata&\nodata&0.2 \nl
$H\alpha/H\beta$&\nodata&\nodata&3.5 \nl
$[OIII]/[OII]$&\nodata&\nodata&18.2 \nl
$[OIII]/[OI]$&\nodata&\nodata&25.6 \nl
$[OI]/H\alpha$&\nodata&\nodata&0.1 \nl 
$[SII]/H\alpha$&\nodata&\nodata&0.2 \nl
$D(4000)_{1\arcsec}$&\nodata&\nodata&$2.15\pm0.18$\nl
$D(4000)_{5\arcsec}$&\nodata&\nodata&$1.96\pm0.14$\nl
\enddata
\end{deluxetable}


\clearpage

\begin{figure}[p]
\plotone{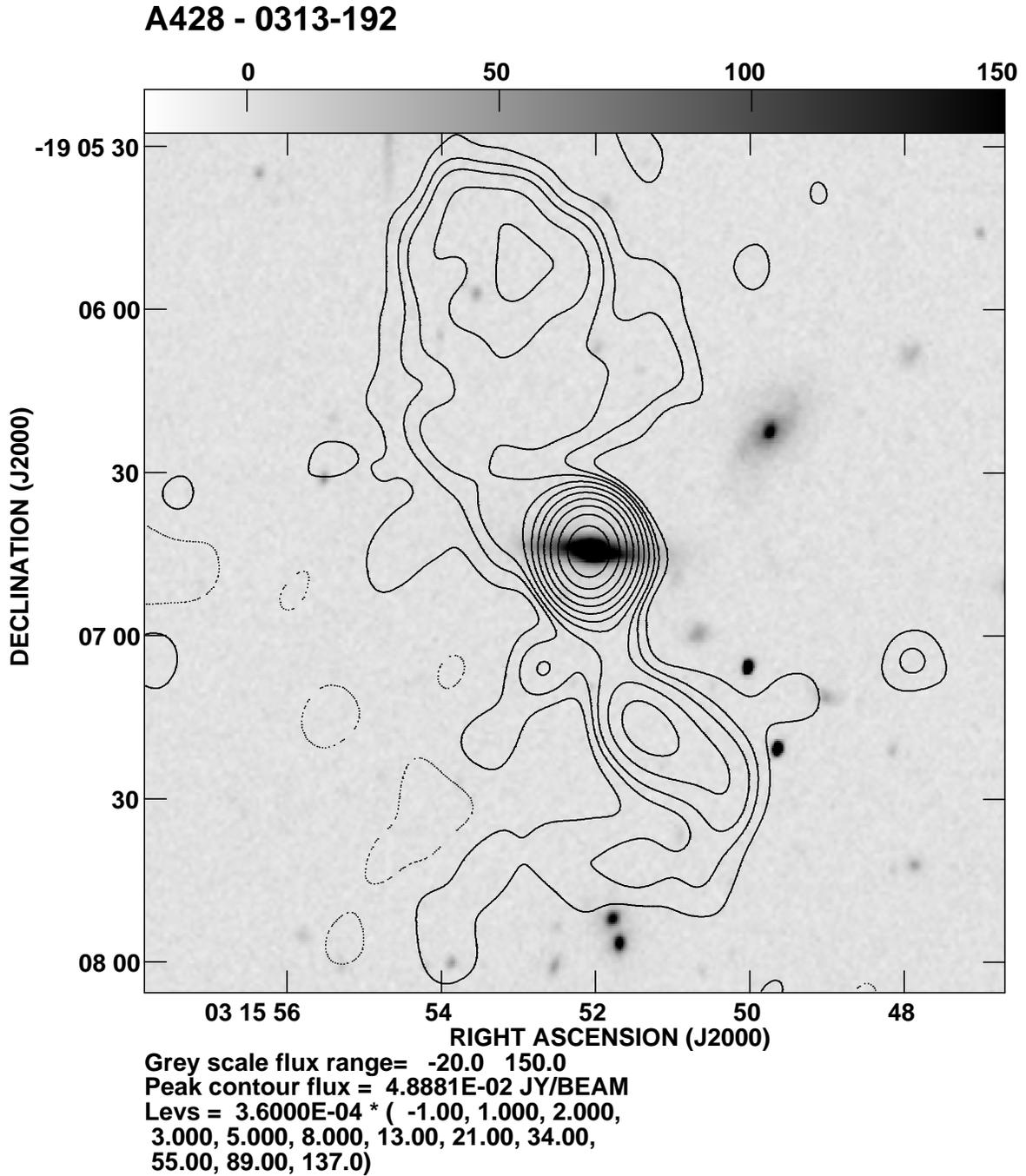}
\caption
{Radio (20cm)/Optical (R-band) overlay for 0313-192.  The 
radio source extends $\approx$ 100 kpc north and south of the host galaxy.} 
\end{figure}

\clearpage

\begin{figure}[p]
\plotone{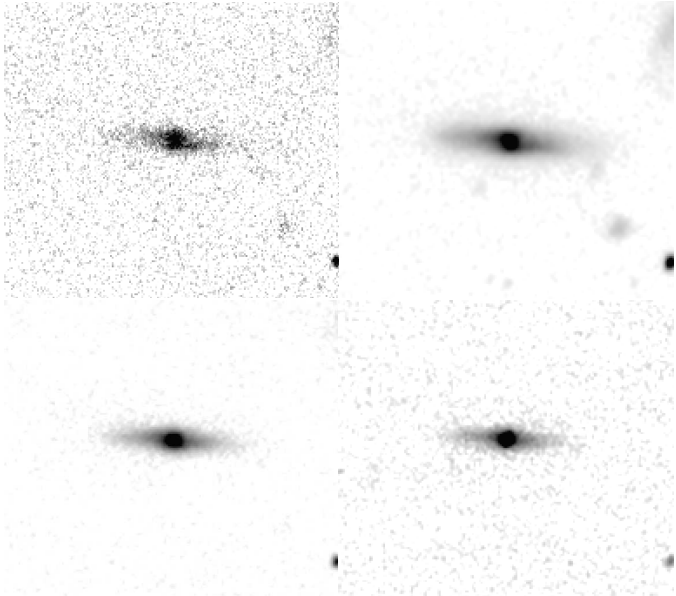}
\caption
{Grey-scale images at B, R, J, and K.  We have applied a 
transfer function scaled by the square-root. 
Upper left=B (KPNO), upper right=R, 
lower left=J, lower right=K.}
\end{figure}

\clearpage

\begin{figure}[p]
\plotfiddle{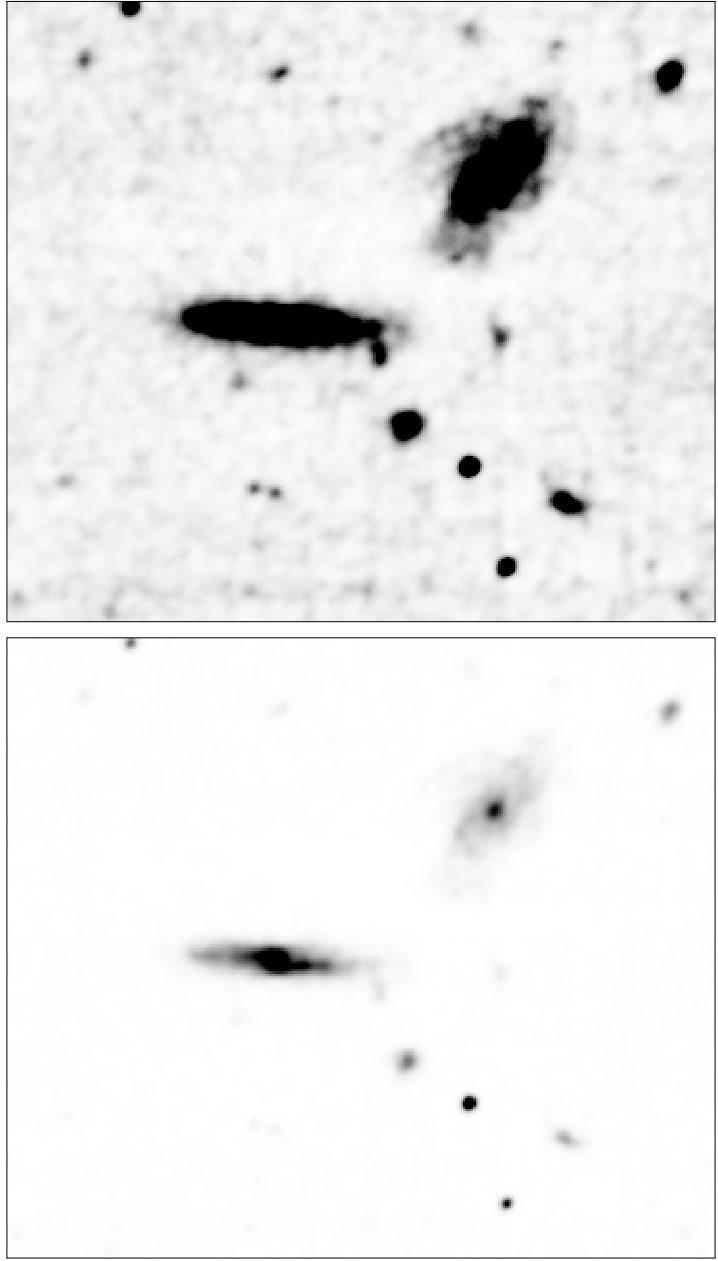}{6in}{0.}{100}{100}{-323}{-135}
\caption
{A deep B-band grey-scale image from the ARC 3.5m telescope.
Top: transfer function chosen to accentuate low surface-brightness features
and to look for evidence of an interaction with the neighboring galaxies. 
Bottom: transfer function chosen to bring out the features in the disk.} 
\end{figure}
  
\clearpage

\begin{figure}[p]
\plotone{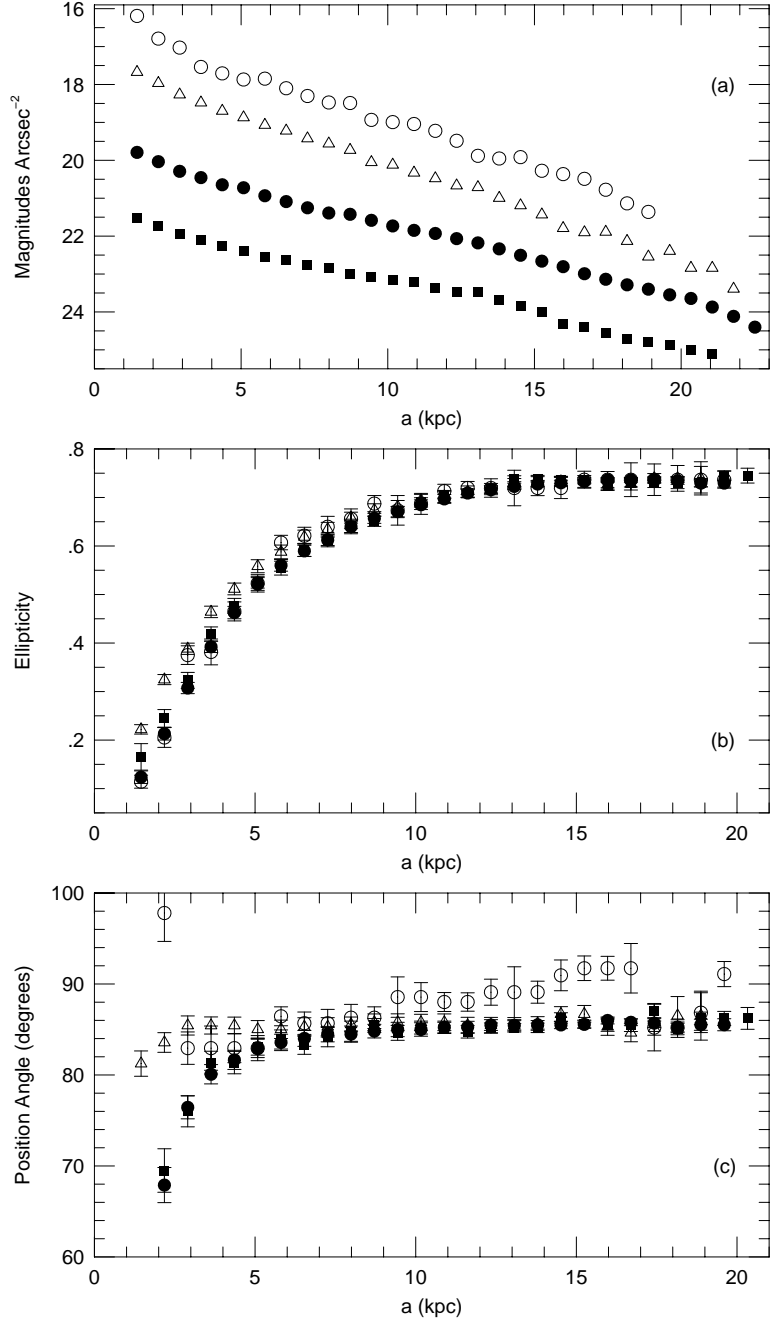}
\caption
{Results of isophotal fitting at B (filled squares), R (filled
circles), J (open triangles), and K (open circles). (a) The surface-brightness
profiles for each color. The surface brightness is corrected to the rest-frame
of the galaxy. (b) The ellipticity profiles as a function of radius at each color.
(c) The variation in position angle with color.}
\end{figure}

\begin{figure}[p]
\plotone{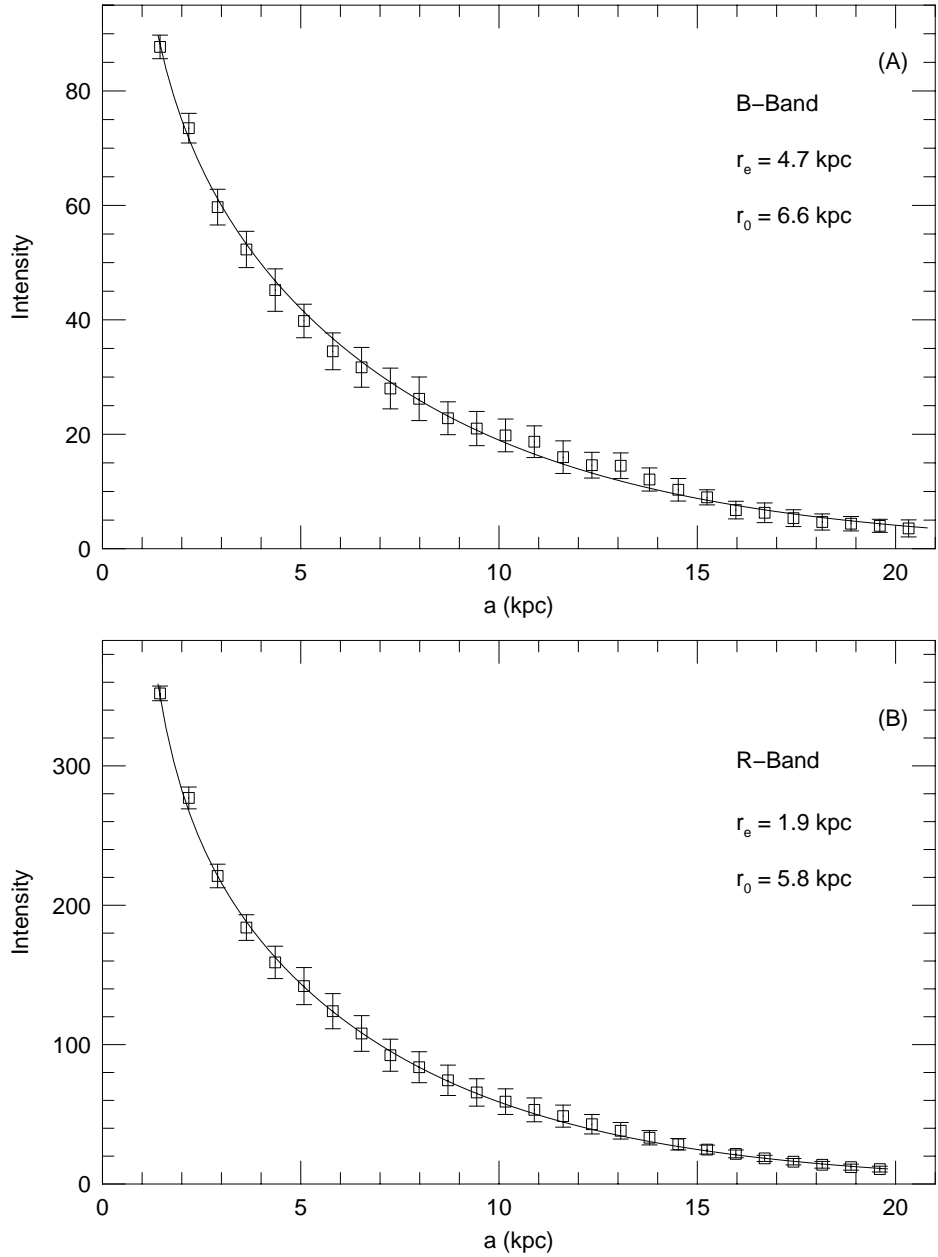}
\caption
{The decomposition of the major-axis intensity profile into 
bulge and disk components.  The solid-line is the best-fit combined bulge+disk model. 
The fit parameters are labeled on the plot.  (a) B-band, (b) R-band.}
\end{figure}

\clearpage

\begin{figure}[p]
\plotone{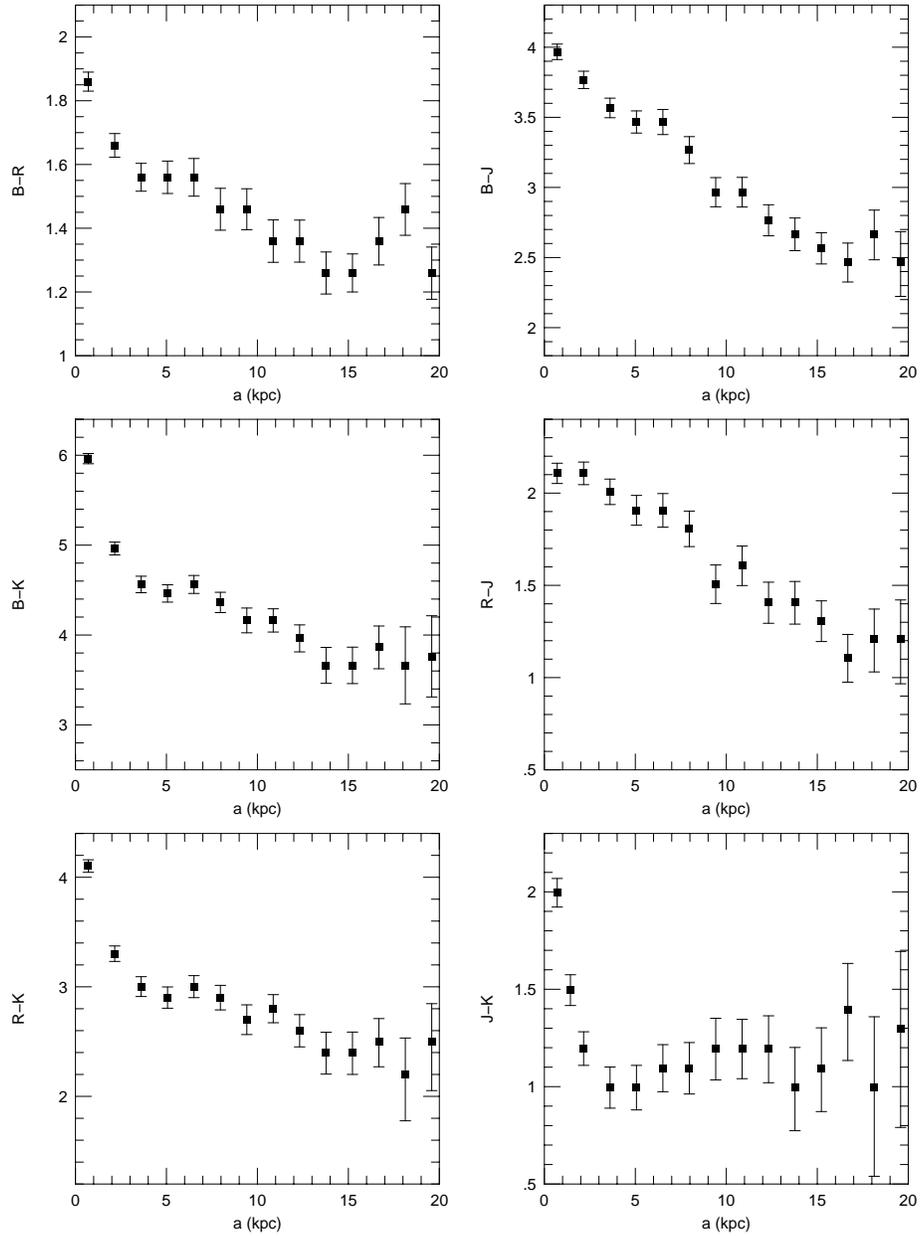}
\caption
{Color profiles for 0313-192.  The colors have been corrected for 
galactic extinction. No K-correction has been applied.  Notice the highly 
reddened core relative to the disk colors.} 
\end{figure}

\clearpage

\begin{figure}[p]
\plotone{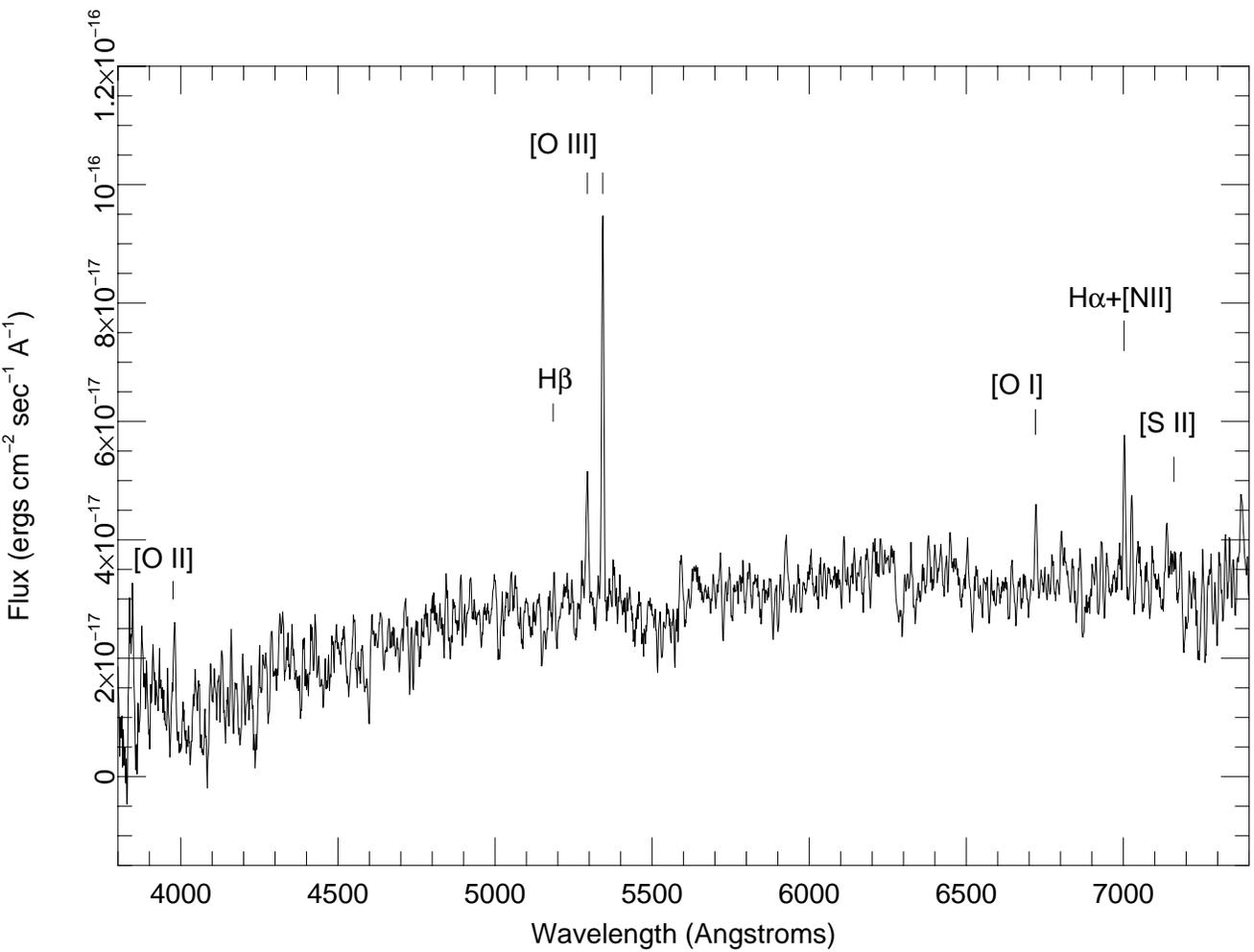}
\caption
{The nuclear optical spectrum of 0313-192.  The slit orientation 
was North-South, perpendicular to the disk.  The salient features are the AGN-like
emission lines, seen through the strong [OIII] and [O I] lines, and relatively 
weak Balmer emission.} 
\end{figure}

\clearpage

\begin{figure}[p]
\plotone{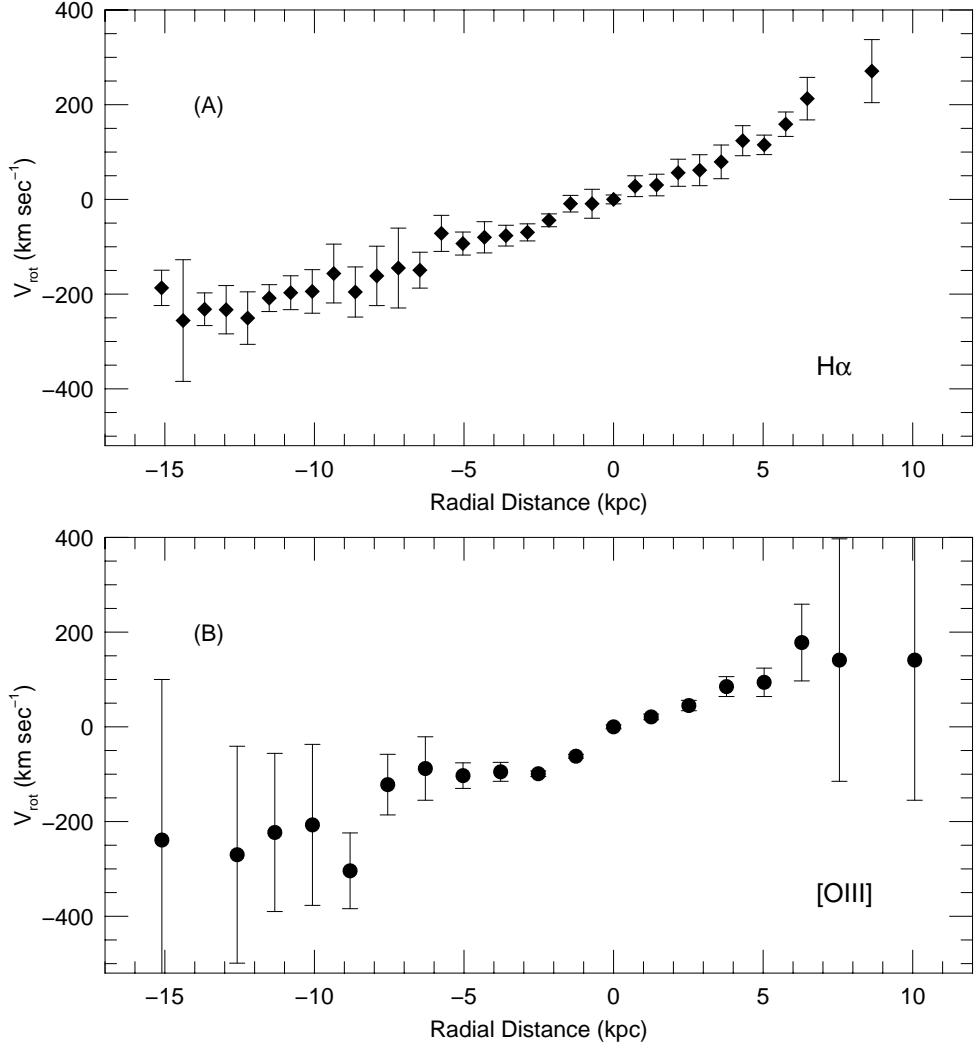}
\caption
{The optical rotation curves measured from (a) $H\alpha+[NII]$, and 
(b) [OIII]$\lambda4959+5007$.  The radial distance is measured relative to the 
nucleus in kpc, with east defined as positive.  We adopt a maximum rotational
velocity of $v_{rot}=208$ km s$^{-1}$.}   
\end{figure}

\end{document}